\documentclass[]{aastex701}

\begin{document}

\title{Ancient relic moderately metal-rich bulge cluster Tonantzintla~2}

 
\correspondingauthor{Beatriz Barbuy}
\email{b.barbuy@iag.usp.br}

\author[0000-0001-7939-5348]{Sergio Ortolani}
\affiliation{Dipartimento di Fisica e Astronomia, Universit\`a di Padova, Vicolo dell'Osservatorio 2, Padova, 35122 Italy}
\affiliation{Centro di Ateneo di Studi e Attivit\`a Spaziali “Giuseppe Colombo”, Via Venezia 15, Padova, 35131 Italy} 
\affiliation{INAF-Osservatorio di Padova, Vicolo dell'Osservatorio 5, Padova, 35122 Italy}
\email{sergio.ortolani@unipd.it}

\author[0000-0001-8052-969X]{Stefano O. Souza}
\affiliation{Max Planck Institute for Astronomy, K\"onigstuhl 17, D-69117 Heidelberg, Germany}
\email{s-souza@mpia.de}

\author[0000-0003-1149-3659]{Domenico Nardiello}
\affiliation{Dipartimento di Fisica e Astronomia, Universit\`a di Padova, Vicolo dell'Osservatorio 2, Padova, 35122 Italy}
\affiliation{INAF-Osservatorio di Padova, Vicolo dell'Osservatorio 5, Padova, 35122 Italy}
\email{domenico.nardiello@unipd.it}

\author[0000-0001-9264-4417]{Beatriz Barbuy}
\affiliation{Universidade de S\~ao Paulo, IAG, Rua do Mat\~ao 1226, Cidade Universit\'aria, S\~ao Paulo 05508-900, Brazil}
\email{b.barbuy@iag.usp.br}

\author[0000-0003-3336-0910]{Eduardo Bica}
\affiliation{Universidade Federal do Rio Grande do Sul, CP 15051, Porto Alegre, 91501-970 Brazil}
\email{00006798@ufrgs.br}

\author[0000-0002-7552-3063]{Bernardo P.L. Ferreira}
\affiliation{Universidade de S\~ao Paulo, IAG, Rua do Mat\~ao 1226, Cidade Universit\'aria, S\~ao Paulo 05508-900, Brazil}
\email{plf.bernardo@gmail.com}

\author[0000-0003-1269-7282]{Cristina Chiappini}
\affiliation{Leibniz Institute for Astrophysics, An der Sternwarte 16, Potsdam, 14482,  Germany}
\email{cristina.chiappini@aip.de}

\author[0000-0003-3526-5052]{Jos\'e G. Fern\'andez-Trincado}
\affiliation{Universidad Cat\'olica del Norte, N\'ucleo UCN en Arqueolog\'ia Gal\'actica - Inst. de Astronom\'ia, Av. Angamos 0610, Antofagasta, Chile}
\email{jose.fernandez@ucn.cl}

\author[0000-0001-6541-1933]{Heitor Ernandes}
\affiliation{Lund Observatory, Department of Geology, Lund University, S\"olvegatan 12, Lund, Sweden}
\email{heitor.ernandes@alumni.usp.br}

\begin{abstract}
The assembly history of the Galactic bulge is intimately tied to the formation of the proto–Milky Way, yet reconstructing this early phase is difficult because mergers and secular evolution have erased most of its original structure. Among present-day stellar systems, only globular clusters retain the ancient signatures needed to trace these primordial building blocks. 
Here we present the most detailed characterization to date of Tonantzintla~2, a prime candidate for a relic of the Milky Way’s primordial bulge. It is a moderately metal-rich globular cluster projected onto the bulge that has remained largely unexplored despite its potential to constrain the early formation of the inner Milky Way. We derive its fundamental parameters using proper motion-corrected \emph{Hubble} Space Telescope WFC3  and ACS photometry. By applying an isochrone fitting to very clean data, we obtain an age of 13.58$^{+0.72}_{-1.0}$ Gyr, a reddening  E(B-V) = 1.44$\pm$0.02, a metallicity  [M/H]=-0.68$^{+0.04}_{-0.05}$, and a heliocentric distance of d$_{\odot}$ = 7.38$^{0.13}_{0.08}$ kpc. A complementary chemical-abundance analysis of seven member stars from APOGEE high-resolution spectroscopy reveals an enrichment pattern consistent with an in-situ origin. Tonantzintla~2 is among the oldest globular clusters studied in the literature,  and the oldest so far analyzed in the Galactic bulge. Its age places a stringent constraint on the onset of the bulge formation, implying that star formation in the inner Galaxy began within $\sim$0.2 Gyr of the Big Bang and that Tonantzintla~2 represents an exceptional relic of the Milky Way's earliest chemical enrichment.

\end{abstract}

\keywords{\uat{Galaxies}{573} --- \uat{Cosmology}{343} --- \uat{High Energy astrophysics}{739} --- \uat{Interstellar medium}{847} --- \uat{Stellar astronomy}{1583} --- \uat{Solar physics}{1476}}


\section{Introduction}
\label{section:introducao}

The moderate-metallicity globular clusters ($-$1.4 $<$ [Fe/H] $ < -$0.2) projected in the direction of the Galactic bulge are ancient systems that should trace the earliest phases of the Milky Way's formation. \citet{bica24} reported an updated census of bulge globular clusters, together with their metallicities and ages. Yet for a significant fraction of the moderately metal-rich ($-$0.8 $<$ [Fe/H] $<$ $-$0.4) clusters, age estimates remain absent. These clusters occupy a crucial regime for establishing the Galactic bulge age-metallicity relation. While the metal-rich end ([Fe/H] $> -$0.4) has been anchored by the detailed analysis of the twin metal-rich globular clusters NGC~6528 and NGC~6553 by 
\citep{ortolani25}, and the moderately metal-poor end ($-$1.4 $ <$ [Fe/H] $ < -$1.0) is constrained by the clusters studied in \citet[][see their Table 3]{souza24}, the intermediate, moderately metal-rich regime has remained essentially unconstrained. With the present study of  Tonantzintla~2, we are now finally able to close this long-standing gap.

Recently, \citet{nepal25} identified a chemically and kinematically distinct, pressure-supported spheroidal bulge component, characterized by a rapid early formation, a high-\(\alpha\) sequence with minimal metallicity dependence, and a metallicity distribution function (MDF) sharply peaked at \([\mathrm{Fe/H}] \approx -0.7\), most probably predating both the thick disc and the bar. The MDF peak they find thus defines the characteristic metallicity of the primordial bulge, offering a new framework for interpreting the origin of stellar systems in this metallicity regime.

In this work we analyse Colour-Magnitude Diagrams (CMDs) of the globular cluster Tonantzintla~2 (hereafter Ton~2), having a metallicity of [Fe/H] $\sim$ $-$0.7
\citep{fernandez-trincado22}, essentially identical to the MDF peak found in \citet{nepal25}. While \citet{nepal25} used field stars and could not directly obtain ages for this spheroidal component, the authors speculate that the identified stellar population may represent the oldest fossil record of the Milky Way’s assembly. By determining a precise age for Ton~2 in
this same metallicity regime, we provide a direct constraint on the epoch at which
the primordial bulge formed and refine the bulge age–metallicity relation in the
critical interval it occupies.

 Ton~2 was discovered by \citet{pismis59} at the Tonantzintla Observatory in Mexico, 
 using Schmidt photographic plates in the red. This allowed to further detect reddened bulge clusters.
\citet{djorgovski93} presented revised coordinates.
It is located at equatorial coordinates ($\alpha$,$\delta$) =
$17^{h}36^{m}10.5^{s}$, -38$^{\circ}$33'12.0" and
Galactic coordinates  
(l,b) = 350.79645$^{\circ}$, $-$03.4233$^{\circ}$.  
\citet{bica96} analysed V,I CMDs of Ton~2, deriving a distance to the Sun of
d$_{\odot}$ = 6.4 kpc, and a reddening of E(B-V) = 1.26.
The unique thorough spectroscopic analysis of member stars carried out,
from APOGEE spectra \citep{majewski17},
by \citet{fernandez-trincado22}, resulted in a metallicity of
[Fe/H] = $-$0.70$\pm$0.05. They also found a large spread of N 
abundances which is also correlated with s-process (Ce) elements. 
Ton~2 shows a loose structure, with a concentration parameter of c = 1.5
 \citep{trager95}.

{Ton~2 has received different orbital classifications in the literature: it has been associated with the main bulge \citep{belokurov24}, with the thick disk \citep{perez-villegas20}, and as a low-energy cluster \citep{massari19} or a possible member of the ancient Kraken accretion event \citep{callingham22}. These conflicting classifications clearly highlights the need for firm age and chemical constraints in order to clarify its origin and connection to the early bulge.

In the present work, we carry out a CMD study of the globular cluster Ton~2, together with a complementary chemical abundance analysis aimed at establishing its origin within the in situ bulge population. In particular, we complement the existing APOGEE abundance measurements for seven cluster members with new determinations of manganese and aluminum, key elements for distinguishing between accreted and in situ stellar populations \citep{das20}}. In Sect. \ref{sec:2} the observations and data reductions are described.
In Sect. \ref{sec:3}  isochrone fitting is applied to CMDs and described. 
In Sect. \ref{sec:4} the results, including chemo-dynamical constraints are discussed.
Concluding remarks are given in Sect. \ref{sec:disc}.



\section{Observations and data reduction}
\label{sec:2}

Our analysis combines \emph{Hubble} Space Telescope 
(\emph{HST}) WFC3 and ACS photometry
with ground-based FORS2/VLT observations, enabling proper-motion cleaning from
two-epoch data and yielding a high-fidelity CMD that isolates the cluster’s evolutionary
sequences.

The \emph{HST} observations were carried out during GO-14074 (PI: Cohen) in WFC3/IR filters F110W and F160W, with a total exposure time of $\sim 1271$~s in each filter, and in ACS/WFC filter F606W ( $1 \time 40$~s$~+~4\times 495$~s). The mean epoch of observations is 2016.48. Ground-based data were obtained with 
the FORST2 spectrograph at the Verly Large Telescope (VLT) of the European Southern Observatory (ESO) during the ESO programme 0113.D-0065 (PI: Monaco). In this work, we only adopted images in R\_SPECIAL filter collected during the night of 5th July, 2024 (epoch 2024.51). 

For \emph{HST} data, we adopted the two-step procedure described in \citet{nardiello18} to obtain the final astro-photometric catalog: all the images are first analysed to extract positions and fluxes with the \texttt{hst1pass} routine and ad-hoc point-spread functions (PSFs) (\citealt{anderson22}) from each image; we derived astrometric and photometric transformations between images and we adopted images, transformations and PSFs to analyse with KS2 (\citealt{anderson08}) all the images simultaneously to obtain deeper photometry and precise positions and magnitudes. We cleaned the final catalog from bad measurements by using the diagnostics output of the data reduction as done in \citet{nardiello18b}. 

FORS2 data were reduced by using the softwares described in  \citet{2015A&A...573A..70N}, that are based on the use of empirical PSFs and neighbour subtraction. We derived the geometric distortion of the two FORS2 CCDs by using the procedure described by \citet{2022MNRAS.515.1841G}.

For Ton~2 we calculated the displacements of the stars between first and second epoch following the method presented in \citet{libralato21} that is based on the use of a network of cluster bona-fide stars that allows local transformations. Using the same philosophy, we corrected the differential reddening that affects the magnitudes of Ton~2 stars as described in \citet{milone12}.

\begin{figure}
\centering
\includegraphics[width=\columnwidth]{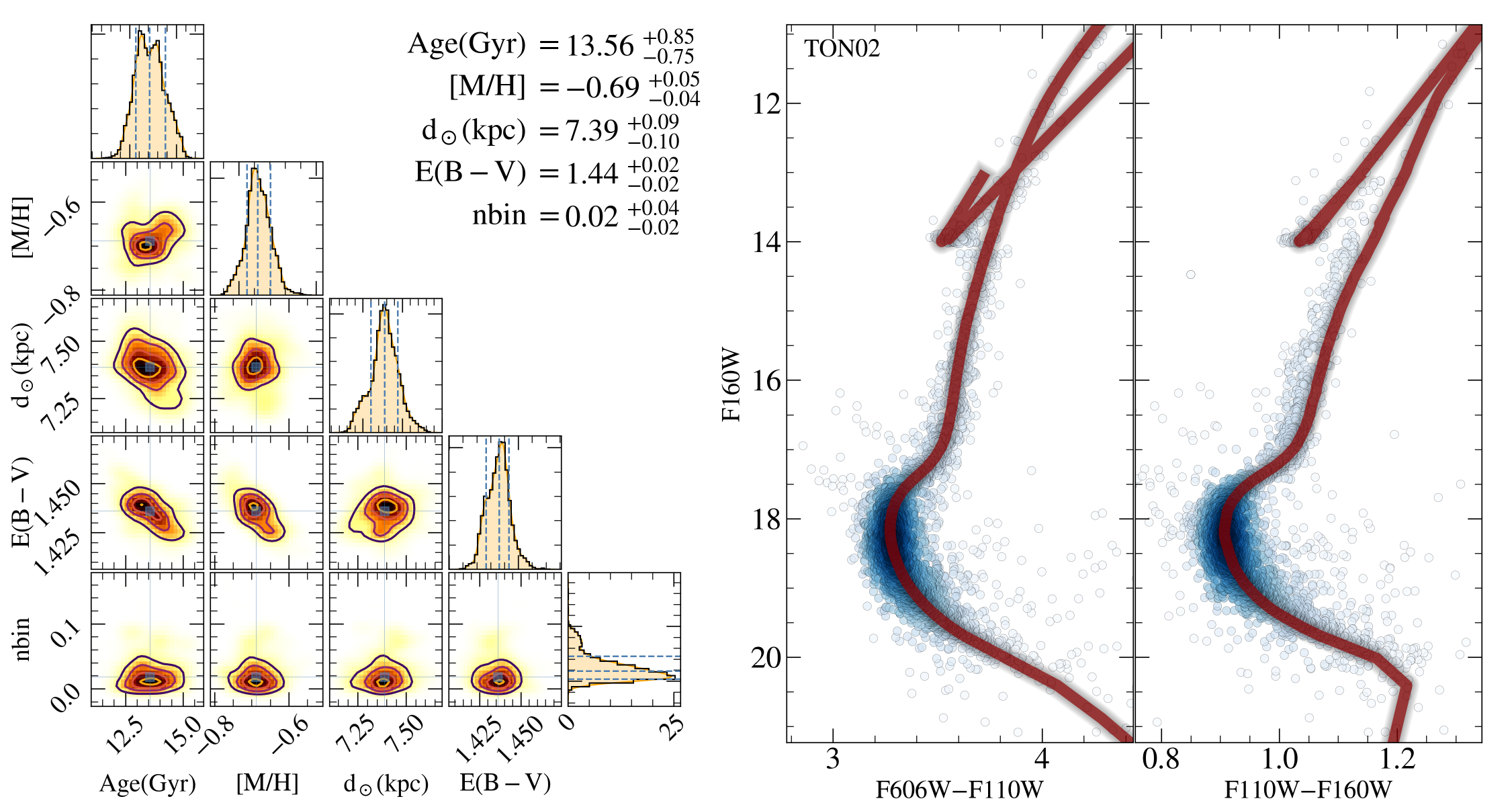}
\caption{Ton~2 F160W vs. F606W-F160W and F160W vs. F110W-F160W CMDs
(right panels), and corner plot of the parameters (left panels).
\textbf{A best fit was found for age of 13.56$^{+0.85}_{-0.75}$ Gyr, 
[M/H]=-0.69$^{+0.05}_{-0.04}$,
d$_{\odot}$ = 7.39$^{+0.09}_{-0.19}$ kpc, and  E(B-V)=1.44$\pm$0.02.} }
\label{fig1}
\end{figure}

\section{Colour-magnitude diagrams and isochrone fitting}
\label{sec:3}

The isochrone fitting procedure, carried out with the code SIRIUS \citep{souza20},
is also well-detailed in \citet{souza23}, \citet{souza24} and \citet{ortolani25}. 
The code enables \textbf{us} to derive  the fundamental parameters 
\textbf{such as} age, metallicity, distance, 
reddening, and total-to-selective extinction ratio (R$_V$).
The PAdova and TRieste Stellar Evolution Code (PARSEC)\footnote{Version 3.8, \url{https://stev.oapd.inaf.it/cgi-bin/cmd_3.8}} \citep{bressan12} isochrones were used to build the synthetic diagrams. More details are given in Appendix \ref{secA1}.

The two proper motion-cleaned  CMDs are shown in  Figure \ref{fig1}, using the ACS wide-field camera, with filter F606W, and
the WFC3 camera with filters F110W and F160W.
The simultaneous fit of F160W vs. F606W-F160W, and F160W vs. F110W-F160W CMDs allows \textbf{us} to derive the proper value of the reddening law R$_{\rm V}$
\citep{pallanca21}.

The resulting combined fit yields an age of 13.58$^{0.72}_{-1.0}$ Gyr,  where
internal uncertainties of the Bayesian method are assumed. 
A reddening of E(B-V) = 1.44$\pm$0.02, a metallicity of [M/H]=$-$0.68, a distance from the Sun of d$_{\odot}$ = 7.38$^{0.13}_{0.08}$ kpc, and R$_{\rm V}=2.9\pm0.1$ are obtained.

 The total-to-selective absorption parameter R$_{\rm V}$ = 2.9 is somewhat lower than the standard value
 of 3.1, which is compatible with the location of the cluster at about $3.4^\circ$ below the Galactic plane.
 As a matter of fact, the extinction maps by \citet{zhang23} indicate similar values
(2.9 $<$  (R$_{\rm V}$ $<$ 3.0) for the symmetric side of the bulge. We note that, from the same method of
 fitting CMDs in the optical and near-infrared, we derived R$_{\rm V}$ = 2.6 for Palomar~6 \citep{souza21}, and R$_{\rm V}$ = 2.84$\pm$0.02  for NGC~6355 \citep{souza23}.
 
 Finally, the age of 13.6 Gyr is the oldest of all analysed bulge clusters so far, and it will be further
 discussed in the age-metallicity subsection below.
 
\section{Discussion}
\label{sec:4}

In this Section we discuss the cluster orbits, chemical composition, and its old age.
In Appendix \ref{secA2}, \textbf{we} report literature ages, reddening E(B-V), distances, and \textbf{of} Ton~2.
For completeness, \citet{harris96} reports a radial velocity of $-$184.4 km.s$^{-1}$.

\subsection{Orbits: a bulge cluster}
The orbits were computed to derive the orbital parameters: 
the apogalactic distance r$_{\rm apo}$,
perigalactic distance r$_{\rm peri}$, 
eccentricity $ecc$,
maximum absolute height relative to the disc $|Z|_{\rm max}$,
the energy E$_{T}$, and the 
angular momentum L$_{Z}$.

For the orbital integration, we adopted
 the analytical approximation given by \citet{sormani22}
for the \citet{portail17} potential,
with the Action-based GAlaxy Modelling Architecture
(AGAMA; \citet{vasiliev19}).
There results:

\par r$_{\rm apo}$: 4.25 ± 0.46
\par r$_{\rm peri}$: 1.03 ± 0.23
\par $ecc$: 0.63 ± 0.08
\par $|Z|_{\rm max}$: 2.19 ± 0.09
\par E$_{T}\times 10^5$: $-$1.09 ± 0.02
\par L$_{Z}\times 10^3$: $-$0.40 ± 0.05

showing that the cluster orbit is contained in the bulge volume.
\textbf{For reference, \citet{nepal25} found that the maximum extension of the spheroidal bulge is defined as $r_{apo}<5$ kpc, $|Z_{max}|<3$ kpc, and higher density for $ecc>0.7$.}

\subsection{Age-metallicity plane: the oldest bulge cluster?}

In Table \ref{tab:ages} we report bulge globular clusters with ages higher than 13.0 Gyr.
As reported in \citet{bica24}, only studies with resolved photometry, and most of them
based on \emph{HST}, or Gemini/GSAOI near-infrared photometry are considered. The
exception is UKS~1 based on the Vista Variables in the Via Lactea (VVV) \citep{minniti10} data.
We also indicate the isochrone set employed, namely, 
Dartmouth or DSED \citep{dotter08}, PARSEC \citep{bressan12}, and BaSTI \citet{pietrinferni06,pietrinferni21}.
\citet{cohen21} used no isochrone sets but derived relative ages by comparing 
CMD fiducial lines of well-known clusters with those of their sample clusters.
As for the ages reported in \citet{bica24}, these same sets of isochrones were used,
besides  the Victoria-Regina code and isochrones \citep{vandenberg14} in a few cases.

\begin{table*}
\small
\caption{\label{tab:ages}
Literature ages of oldest bulge clusters and isochrone sets employed. References: 1. \citet{deras23}; 2. \citet{souza23}; 3. \citet{cohen21}; 4. \citet{pallanca21};
5. \citet{oliveira20}; 6. \citet{fernandez-trincado20};  7. \citet{massari25}.}
\begin{tabular}{lrrrrrrrrrrrrrrr}
\hline
\noalign{\smallskip}
Cluster & \hbox{[Fe/H]} & Age (Gyr)  & reference & \hbox{Isochrone}  \\
\noalign{\smallskip}
\hline
\noalign{\smallskip}
NGC~6316 & $-$0.92$\pm$0.07 & 13.4$^{+0.6}_{-0.5}$ & 1 & PARSEC \\
NGC~6355 & $-$1.39$\pm$0.08 & 13.2$\pm$1.1 & 2 & DSED \\
NGC~6401 & $-$1.15$\pm$0.20 & 13.2$\pm$1.2 & 3 & relative ages \\
NGC~6440 &  $-$0.5$\pm$0.03 & 13.0$\pm$1.5 & 4 & DSED/PARSEC \\
NGC~6453 & $-$1.48$\pm$0.14 & 13.3$\pm$0.8 & 3 & relative ages \\
NGC~6717 & $-$1.17$\pm$0.09 & 13.5$^{+0.67}_{0.76}$ & 5 & DSED/BaSTI  \\
UKS~1 &    $-$0.98$\pm$0.11 & 13.1$^{+0.93}_{-1.29}$ & 6 & DSED \\
ESO 452-11 & $-$0.80$^{+0.08}_{-0.11}$ & 13.59$^{+0.48}_{-0.69}$ &  7 & BaSTI \\
\noalign{\smallskip}
\noalign{\hrule\vskip 0.1cm}
\noalign{\smallskip}                
\end{tabular}
\end{table*}

\begin{figure}
    \centering
    \includegraphics[width=\linewidth]{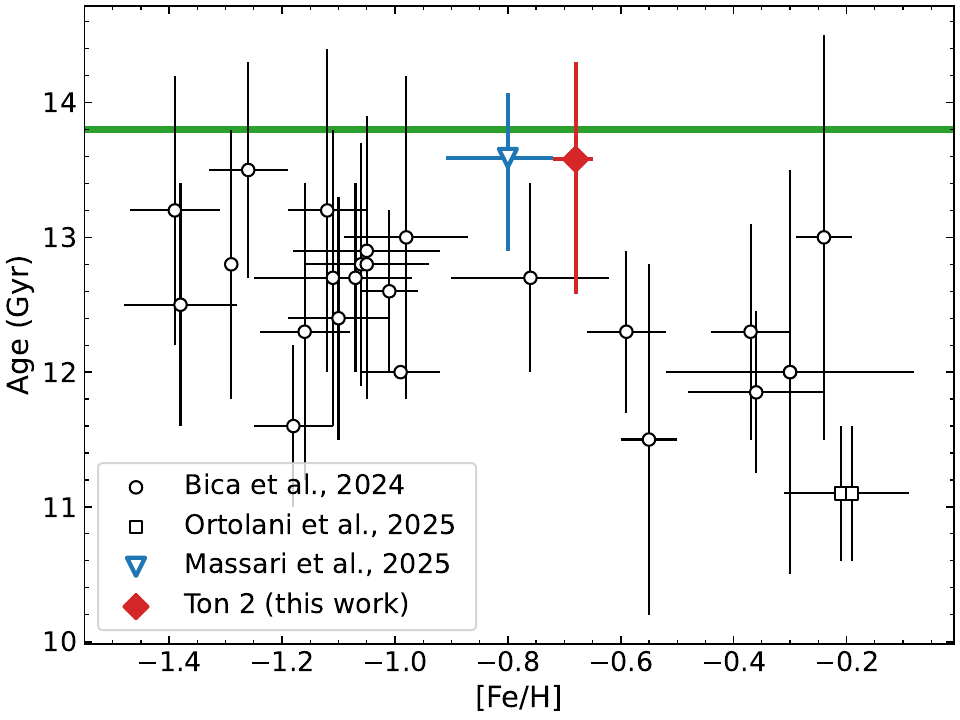}
    \caption{Age vs. metallicity plane for bulge-volume clusters, highlighting the location of Ton~2. 
The values are from \citep{bica24,ortolani25,massari25} . 
The green line represents the age of the universe of 13.801$\pm$0.024
Gyr according to \citet{planck20}, with width showing the uncertainty.}
    \label{fig:AMR}
\end{figure}

Figure \ref{fig:AMR} \textbf{shows} the metallicities and ages of bulge
globular clusters reported in \citet{bica24},  the twin metal-rich clusters
by \citet{ortolani25}, and the location of Ton~2. We also 
show the cluster ESO 452-11 to which \citet{massari25} assigns an age of 13.59$^{+0.48}_{-0.69}$ Gyr.

The position of Ton~2 in the age-metallicity plane for bulge clusters is striking.
Ton~2 appears to have formed very early during the Galaxy formation. It should be a
representative of the proto-bulge. Ton~2 appears to be older than all clusters,
with age only comparable to that of ESO 452-11, another bulge globular cluster 
with similar metallicity and therefore another potential bulge relic (although its in situ 
origin should still be demonstrated via a detailed chemical analysis, similar to that 
present here for Ton 2 in the next section). Although \citet{massari25} report an age of
13.59$^{+0.48}_{-0.69}$ Gyr, the use of BaSTI isochrones \citep{pietrinferni06,pietrinferni21} 
leads to systematically older ages as discussed in \citet{kerber18,gontcharov19,gontcharov20}. 
In Appendix \ref{secA3} we show that from a comparison of BaSTI \citep{pietrinferni06,pietrinferni21}, DSED \citep{dotter08},
and PARSEC \citep{bressan12}, results an age difference of 0.4 dex higher for BaSTI relative to the other models. In addition, it would be important to know the chemical
properties of ESO 452-11, to confirm its in situ origin.

\subsection{Chemical abundances: an in situ bulge cluster}

\begin{table*}
\small
\caption{\label{tab:abon}
Metallicities and chemical abundances for Ton~2,
from \citet{fernandez-trincado22}, C, N, O from \citet{barbuy25b}, and Na, Al, P and Mn computed in this work (referred to as tw).}
\begin{tabular}{l@{}r@{}r@{}r@{}r@{}r@{}cc@{}r@{}r@{}rrrrrr}
\hline
\noalign{\smallskip}
star & \hbox{[Fe/H]} &\hbox{\phantom{-}[C/Fe]} & 
\hbox{\phantom{-}[N/Fe]} & \hbox{\phantom{-}[O/Fe]}  
& \hbox{\phantom{-}[Na/Fe]} & \hbox{\phantom{-}[Mg/Fe]} 
&\hbox{\phantom{-}[Al/Fe]} 
 & \hbox{\phantom{-}[P/Fe]} & \hbox{\phantom{-}[Mn/Fe]} & \hbox{\phantom{-}[Ni/Fe]}\\
 &  & B25b & B25b & B25b  & tw & FT22  & tw/FT22 
 & tw & tw & FT22 \\
\noalign{\smallskip}
\hline
\noalign{\smallskip}
Ton2-2114 & -0.73 & +0.10 & +1.10 & +0.85 & +0.60 & +0.23 & +0.50/+0.39 & +0.39 & -0.08 & +0.09 \\
Ton2-3304 & -0.74 & +0.25 & +0.95 & +0.80 & +0.70 & +0.33 & +0.42/+0.42 & +0.20 & -0.12 & +0.10 \\
Ton2-3312 & -0.55 & +0.25 & +0.50 & +0.90 & +0.15 & +0.38 & +0.32/+0.29 &  ---  & -0.35 & -0.07 \\
Ton2-4199 & -0.77 & +0.20 & +0.80 & +0.70 & +0.75 & +0.22 & +0.49/+0.42 &  ---  & +0.00 & +0.15 \\ 
Ton2-4336 & -0.69 & +0.00 & +1.10 & +0.70 & +0.70 & +0.31 & +0.52/+0.46 &  ---  & -0.10 & +0.09 \\
Ton2-4371 & -0.65 & +0.20 & +0.34 & +0.85 & +0.25 & +0.22 & +0.00/+0.36 & +0.20 &-0.25 & +0.01 \\
Ton2-5151 & -0.75 & +0.10 & +0.65 & +0.55 & +0.50 & +0.30 & +0.04/--- & +0.30 &-0.25 & +0.09 \\
\noalign{\smallskip}
\noalign{\hrule\vskip 0.1cm}
\noalign{\smallskip}                
\end{tabular}
\end{table*}

The unique spectroscopic analysis of member stars of Ton~2 was carried out by \citet{fernandez-trincado22}.
The spectra were obtained within the Apache Point Observatory Galactic Evolution
Experiment II survey (APOGEE-2), obtained as part of the bulge Cluster APOgee Survey 
\citep[CAPOS;][]{geisler25}.
The 7 stars analysed resulted in a mean metallicity of [Fe/H]=$-$0.70, and alpha-element enhancements, with mean values
[Mg/Fe]=+0.29, [Si/Fe]=+0.33, [Ca/Fe]=+0.27. The alpha-like elements Al and Ti also show enhancements of
[Al/Fe]=+0.39, and [Ti/Fe]=+0.29. N, and O are also significantly enhanced.
Finally, [Ce/Fe] shows some spread.

Table \ref{tab:abon} reports the chemical abundances
derived in \citet{fernandez-trincado22}, together with C, N, O, and P analysed in \citet{barbuy25b}, 
a remeasure of aluminum, as well as 
sodium and manganese (Mn) measured in the present work (see Appendix \ref{secA4}).

Figure \ref{insitu} (upper panel) shows [Mg/Mn] vs. [Al/Fe] \citep{das20}  and 
[Ni/Fe] vs. [(C+N)/O] (lower panel) \citep{montalban2021,ortigoza-urdaneta23}
used to discriminate between in situ and ex situ origin.
The location of the 7 studied stars is compared with that of the Reduced Proper Motion (RPM) sample of bonafide inner-galaxy stars from APOGEE \citep{queiroz21},
the accreted structures Gaia-Enceladus-Sausage (GSE) \citep{limberg22}, and Heracles \citep{horta21}.
In the upper panel it is clear that the abundances of 7 stars in Ton~2, given in Table \ref{tab:abon}, 
point toward an in situ origin of the cluster. In the lower panel, only one star shows
[Ni/Fe] slightly negative, whereas all other 6 stars are enriched in Ni, again confirming the in situ origin of the cluster.

\begin{figure}
    \centering
        \includegraphics[width=\linewidth]{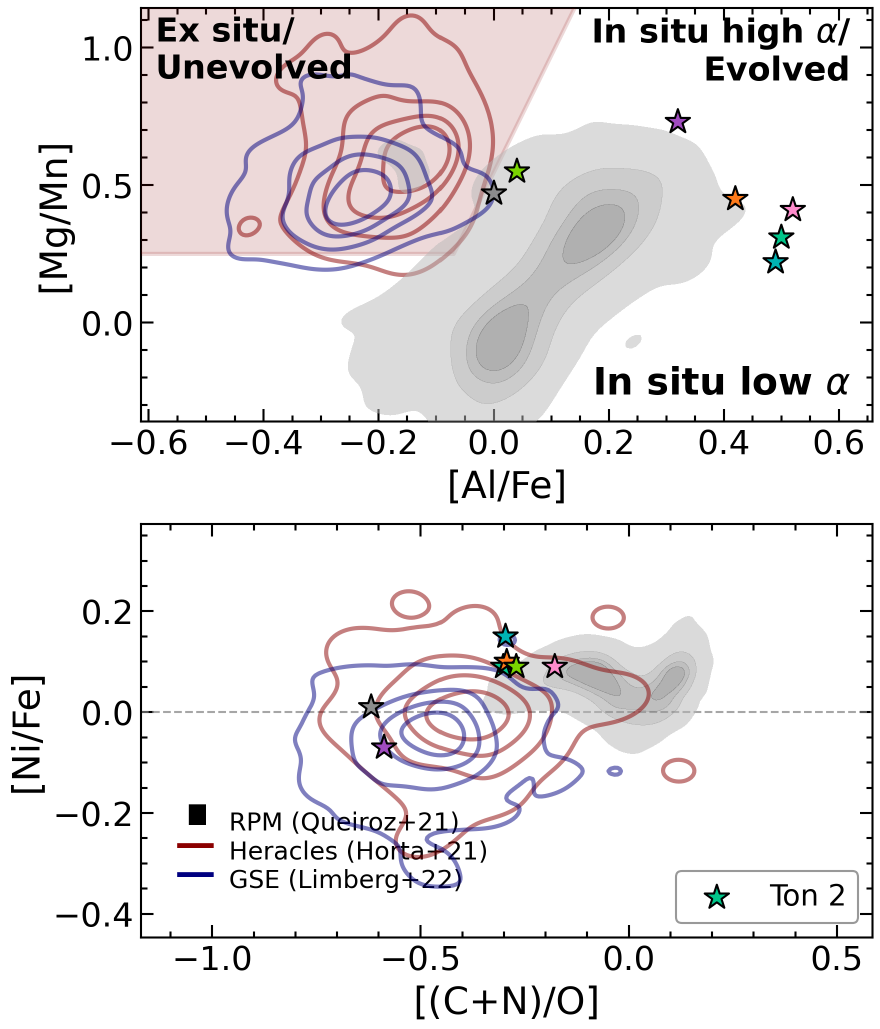}
    \caption{[Mg/Mn] vs. [Al/Fe] (upper panel)  and [Ni/Fe] vs. [(C+N)/O].
    Symbols: black-filled stars: star-members of Ton~2; brown lines: Heracles; dark blue lines: GSE.
    }
    \label{insitu}
\end{figure}

Another interesting further study of these APOGEE spectra was the detection of P-enrichment
in part of the sample stars, for the clusters Tonantzintla~1 and
NGC~6316, whereas for Ton~2 none in 4 stars has shown this effect
\citep{barbuy25b}, and for 3 stars there is noise in the PI line region. In that work,  we confirmed
N and O enhancements, and also a correlation between [P/Fe] vs. [N/O] in Tonantzintla~1
and NGC~6316. For Ton~2 we see a CNO-enhancement, likewise for these 2 other clusters. We note that a NLTE Al abundance would be about 0.1dex
lower than the LTE value.
A final comment is that some of the 7 stars showing high N, O, and Na
could be second generation stars \citep[see][]{bastian18} of this cluster \citep[e.g.][]{fernandez-trincado20}.

Another important indicator is the rather high Al abundance, that points towards an
in situ origin of these clusters \citep{meszaros20}. This contradicts the suggestion by 
\citet{callingham22}, that ranked Ton~2 as having origin in the Kraken structure,
that otherwise was not included as a Kraken member in
\citet{kruijssen20}.

\section{Conclusions}
\label{sec:disc}
Our multi-band \emph{HST} photometry and high-resolution chemical analysis provide the most precise age derivation to date of the bulge globular cluster Tonantzintla-2. We derived a distance of 
 7.38$^{+0.13}_{-0.08}$ kpc, and showed that its orbit is fully confined within the Galactic bulge. Isochrone fitting of proper-motion-cleaned colour-magnitude diagrams yields an age of 13.58$^{+0.72}_{-1.0}$ Gyr, placing Ton~2 among the oldest stellar systems known in the Milky Way, and the oldest so
 far in the Galactic bulge. At its metallicity of [Fe/H] $= -$0.7, Ton 2 sits exactly at the peak of the spheroidal, and high-alpha stellar population, a primordial bulge component recently identified from field stars by \citet{nepal25}, providing a direct age measurement for this population. A detailed high-resolution chemical abundance analysis of seven confirmed member stars, for which we determined Manganese abundances - an element crucial in helping distinguishing in situ from accreted stellar populations, complements our analysis. We find that the chemical abundance pattern -- enhanced $\alpha$-elements, high Al, N and Ni and elevated [Mg/Mn] -- is fully consistent with an in situ origin and incompatible with an accreted structure origin such as Kraken.

These results show that the bulge reached the moderate metallicity of [Fe/H] $\simeq -$0.7 extremely early, within the 200 million years from the Big Bang \citep{planck20}. Ton~2 therefore represents a fossil remnant of the proto-bulge. Its properties (age and metallicity) remind those of bulges and proto-globular clusters, as N-emitters now observed by JWST beyond redshift six \citep{morel25}. This suggests that the Milky Way experienced a similar rapid, early bulge-building phase, linking local relics such as Ton~2 with the processes driving galaxy assembly and chemical enrichment very early on in the history of the Universe. 

 Our work therefore provides, for the first time, the chemical and chronological framework that links this population of very old moderately metal-poor bulge clusters to the metallicity peak ([Fe/H] $\simeq -$0.7) of the primordial bulge. Ton 2 becomes the oldest globular cluster for which this connection can be demonstrated directly.

\begin{acknowledgments}
BB and EB acknowledge partial financial support from FAPESP, CNPq, and CAPES - Financial code 001. CC acknowledge partial financial support from FAPESP.
 SOS acknowledges the DGAPA–PAPIIT grant IA103224 and the support from Dr. Nadine Neumayer's Lise Meitner grant from the Max Planck Society.
 J.G.F-T gratefully acknowledges the grants support provided by ANID Fondecyt Postdoc No. 3230001 (Sponsoring researcher), the Joint Committee ESO-Government of Chile under the agreement 2023 ORP 062/2023, and the support of the Doctoral Program in Artificial Intelligence, DISC-UCN.
 This study was financed, in part, by the São Paulo Research Foundation (FAPESP), Brazil; Process Number 2025/05050-3.
 This research is based on observations made with the NASA/ESA \emph{Hubble} Space Telescope obtained from the Space Telescope Science Institute, which is operated by the Association of Universities for Research in Astronomy, Inc., under NASA contract NAS 5–26555.  
 The HST observations are associated with programme GO-14074 (PI: Cohen).
 \textbf{The data presented in this article were obtained from the Mikulski Archive for Space Telescopes (MAST) at the Space Telescope Science Institute. The specific observations analyzed can be accessed via \dataset[doi:  10.17909/t9-na33-8504]{https://doi.org/10.17909/t9-na33-8504}}.
 Ground-based data were obtained at the FORS2@VLT spectrograph, during the European Southern Observatory 
 programme 0113.D-0065 (PI: Monaco). Funding for the Sloan Digital Sky Survey IV has been provided by the Alfred P. Sloan Foundation, the U.S. Department of Energy Office of Science, and the Participating Institutions. SDSS-IV acknowledges support and resources from the Center for High-Performance Computing at the University of Utah. The SDSS website is www.sdss.org.
\end{acknowledgments}

\begin{contribution}

All authors contributed equally to this project.


\end{contribution}

%



\appendix

\section{The code SIRIUS}\label{secA1}

\textbf{The code SIRIUS \citep{souza20}, employed here, to derive the age of Tonantzintla~2,
adopts} the Bayesian  Markov chain Monte Carlo (McMC) method to obtain 
probability distributions for each parameter by comparing the observed CMDs with synthetic
CMDs built from each set of parameters randomly, during the fitting process. 
The McMC was applied using the \texttt{Python} library, \texttt{emcee} \citep{emcee}, and \texttt{PyDE},\footnote{\url{https://github.com/hpparvi/PyDE}} a global optimisation that uses differential evolution.
The PARSEC isochrones are adopted \citep{bressan12}.
The parameter space spans ages between 7 Gyr and 14 Gyr with intervals of 0.1 Gyr, and metallicities 
between $-2.0$ and $+0.3$ with intervals of $0.05$ dex. For each McMC 
\textbf{realization,} interpolations in the
isochrone grid are carried out, assuming the initial mass function from \citet{kroupa01}. 
The synthetic CMDs 
use magnitudes from the PARSEC isochrone grid according to the parameters selected at each McMC step. 
A fraction of unresolved binaries is adopted. The secondary components are  assigned using a randomly 
selected mass ratio, and their combined magnitudes are computed by summing their fluxes. 
A magnitude-dependent error function derived from the observed photometric errors is applied, 
at the position of the turn-off to preserve alignment with the observed CMD, simulating observational uncertainties.
The extinction correction is computed at each iteration, applying the extinction law of \citet{cardelli89} for a
determined R$_V$ value. The observed distribution of stars in magnitude space results from luminosity function applied to the synthetic sample.

\section{Previous literature on Tonantzintla~2}\label{secA2}

Literature values of reddening of 1.26 $<$ E(B-V) $<$ 1.23 were adopted by 
\citet[][H96]{harris96}, \citet[][B96]{bica96}, \citet[][DB00]{dutrabica00},
\citet[][V18]{vasquez18}, \citet[][G23]{geisler23}, whereas E(B-V) = 1.41$^{+0.5}_{-0.1}$
was given by \citet[][SF11]{schlafly11}, the latter coinciding with our resulting value
of E(B-V)=1.44$\pm$0.02. We note that there is a strong reddening variation across the 
field. This explains why in Figure \ref{fig1} we see two peaks in E(B-V) and in age.
 
\citet[][V21]{vasiliev21} estimate a distance to the Sun of d$_{\odot}$ = 5.9$\pm$0.6 computed from
Gaia data, whereas \citet[][B21]{baumgardt21} gives  d$_{\odot}$ = 6.987$\pm$0.34 
based on a combination of several measurements, including Gaia data.
We find a slightly larger distance of d$_{\odot}$ = 7.38$^{+0.13}_{-0.08}$,
a distance closer to the Galactic center, and clearly in the bulge volume.

\citet[][V18]{vasquez18} and \citet[][G23]{geisler23} derived metallicities of
$-$0.73$\pm$0.13, and $-$0.61$\pm$0.03 respectively from the CaII Triplet lines (CaT),
whereas a thorough spectroscopic analysis by \citet[][F-T22]{fernandez-trincado22}
resulted in a metallicity of [Fe/H] = $-$0.70$\pm$0.05. Our derived value of
[M/H] = $-$0.68$^{+0.04}_{-0.08}$ is somewhat lower than that of F-T22, given that
it takes into account the alpha-enhancement around [$\alpha$/Fe] $= +0.4$ (F-T22).

\section{Age vs. different isochrone sets}\label{secA3}

\begin{figure}
    \centering
    \includegraphics[width=\linewidth]{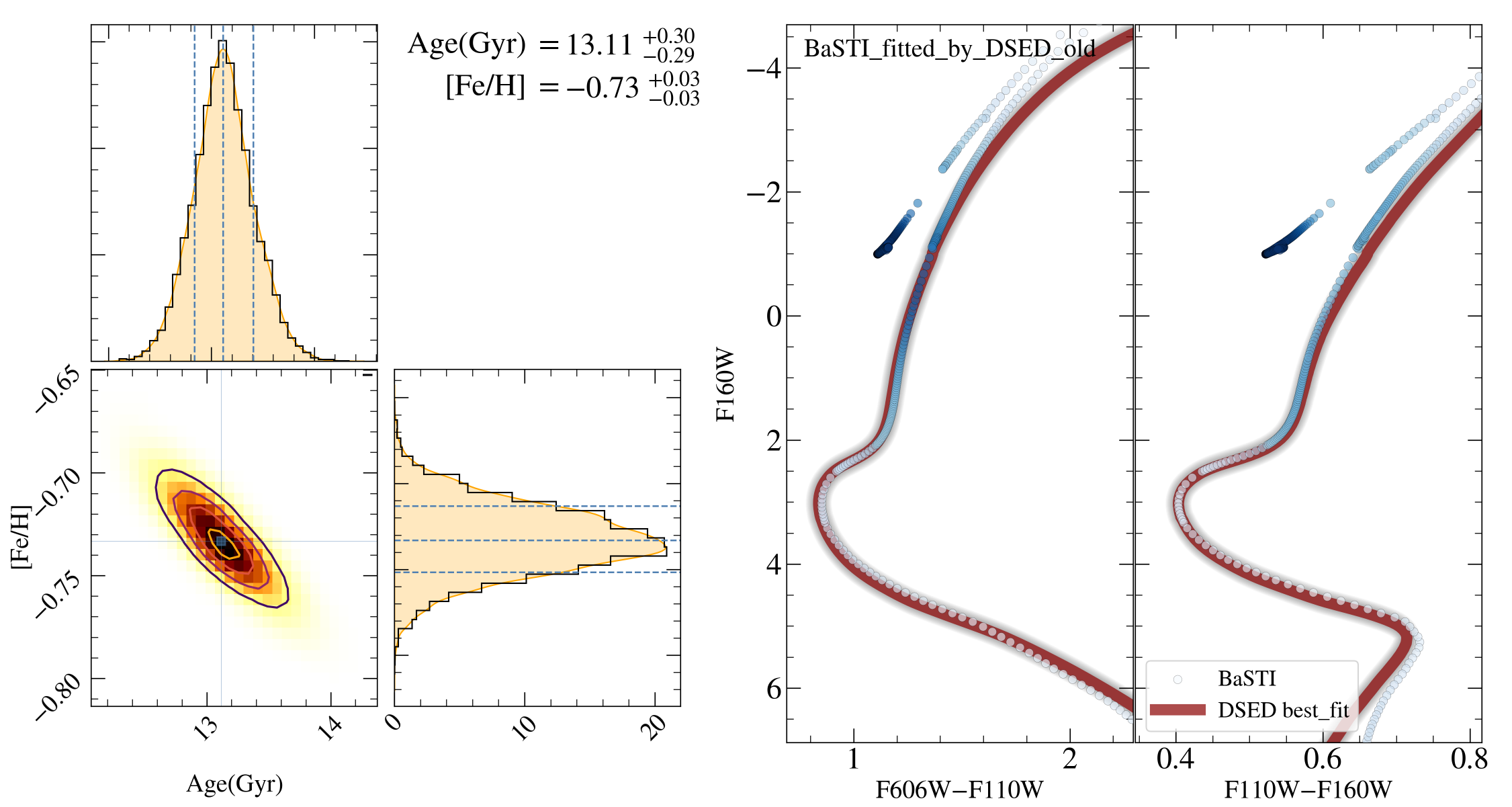}
        \includegraphics[width=\linewidth]{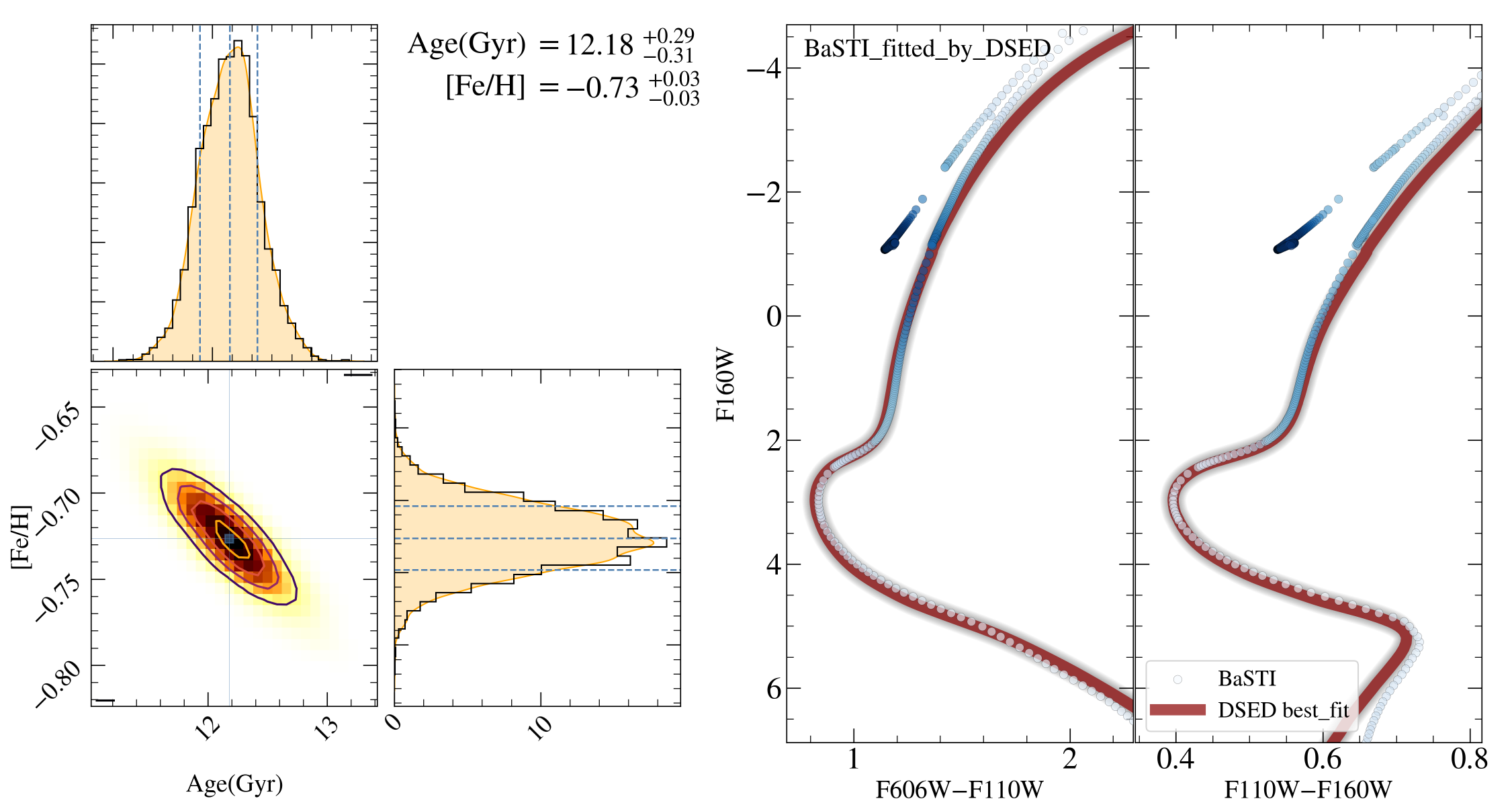}
    \caption{BaSTI isochrones of 13.5 and 12.5 Gyr, fitted with DSED isochrones,
    and corresponding age.}
    \label{fig:basti}
\end{figure}

For illustrating the difference in age as derived from different isochrone sets,  in Figure \ref {fig:basti} we carried out a
comparison of BaSTI isochones \citep{pietrinferni06} of ages 13.5 and 12.5 Gyr and [Fe/H] = -0.7
fitted by DSED isochrones \citep{dotter08}, giving 13.11$^{+0.30}_{-0.28}$ and 12.17$^{+0.29}_{-0.31}$ respectively;
For Ton2’s metallcity ([Fe/H]=-0.70), there is a very good match between BaSTI and DSED in the 
main-sequence turn-ofhf and subgiant branch (MSTO+SGB), but for a 0.4 Gyr younger DSED isochrone,
representing a systematic difference.

For a comparison with PARSEC \citet{bressan12} isochrones, resulting in:
1-The red giant branch (RGB) tilting is the largest difference. BaSTI is steeper, eventually in 
better agreement with our data, but this is not the case near the subgiant branch (SGB) base.
A younger BaSTI gives the best fit to PARSEC;
2-Turn-off (TO) shows a shift. It is not obvious if BaSTI gives an older or younger age. 
Shifting and forcing the fit to PARSEC eventually will lead to a best fit  with 13.5 Gyr;
3-The largest discrepancy is the horizontal branch (HB). BaSTI is too faint, and an older age
is needed to fit simultaneously HB and TO. Consequently, since \citet{massari25} used the HB fit, 
they found an older age relative to PARSEC.
Finally, the data for ESO 452-11 are not clean from field stars, therefore the age derivation is uncertain.

\section{Manganese abundances }\label{secA4}

In order to analyse indicators of stellar populations ex situ or in situ, we derived Mn abundances. We also rederived Al, and Na abundances.
The code Turbospectrum for spectrum synthesis is used \citep{plez12}.

\begin{figure}
    \centering
    \includegraphics[width=\linewidth]{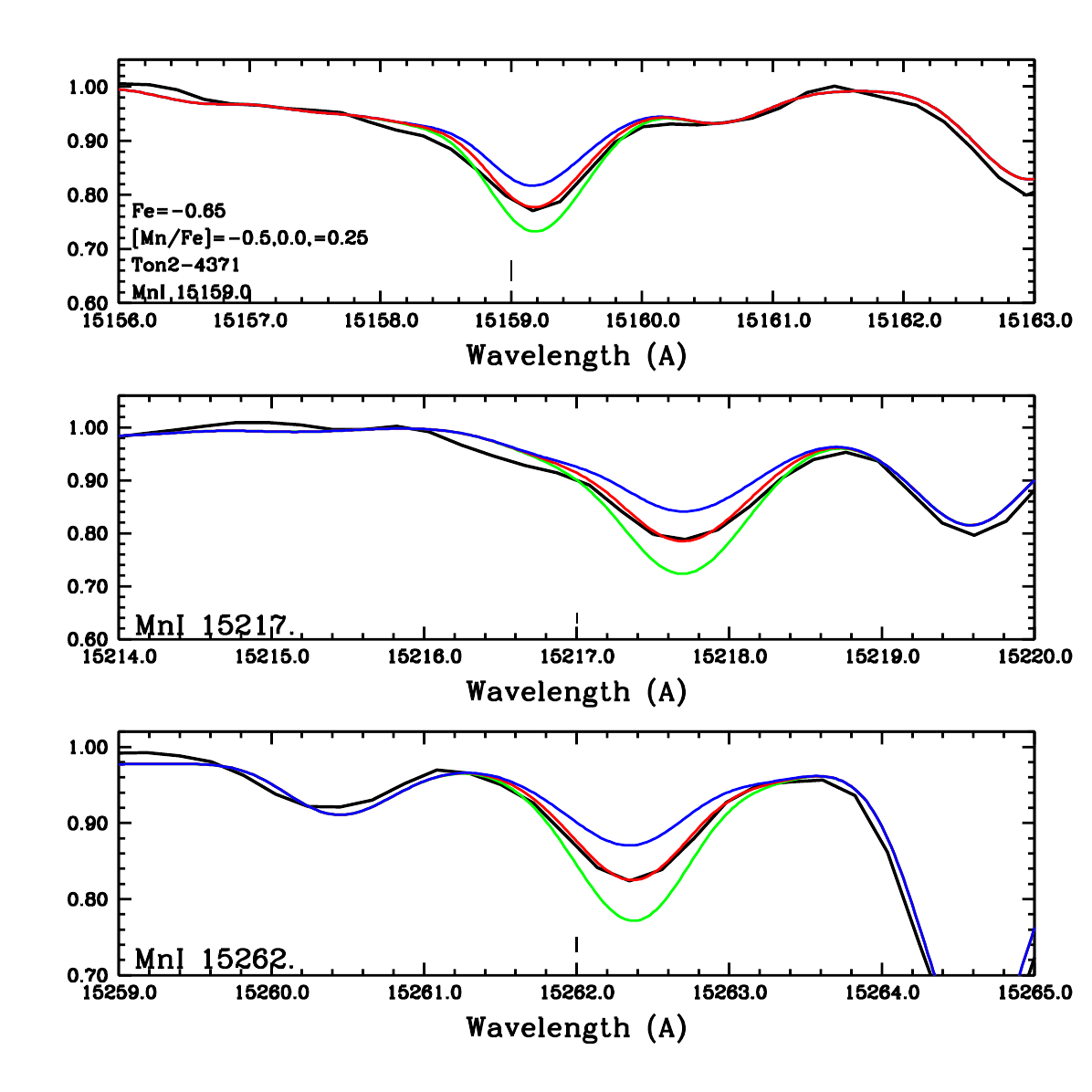}
    \caption{Fit to Mn lines for star Ton2: 2M17361421-383431.}
    \label{fig:mn}
\end{figure}

Figure \ref{fig:mn} shows the fit to the three Mn lines MnI 15159.200,  15217.793, 15262.702 {\rm \AA}
 in star Ton2:  2M17361421-3834371,
illustrating the good quality of the fit, and therefore of the Mn abundances. We also rederived the abundances of Na and Al,
from the lines NaI 16388.85 {\rm \AA}, and
AlI 16718.957, 16750.539 and 16763.359 {\rm \AA} lines. 

\bibliography{sn-bibliography}{}
\bibliographystyle{aasjournalv7}



\end{document}